\documentclass[11pt]{article}

\usepackage[final]{acl}

\usepackage{times}
\usepackage{latexsym}
\usepackage{amsmath} 
\usepackage{amssymb}
\usepackage{tikz}
\newcommand*\circled[1]{\tikz[baseline=(char.base)]{
    \node[shape=circle,draw,inner sep=0.5pt] (char) {\scriptsize #1};}}
\usepackage{array}
\usepackage{booktabs}
\usepackage{multirow}
\usepackage[most]{tcolorbox}
\usepackage{cuted}

\usepackage{algorithm}
\usepackage{algorithmic}

\usepackage{enumitem}
\usepackage{arydshln}
\usepackage[T1]{fontenc}

\usepackage[utf8]{inputenc}

\usepackage{microtype}

\usepackage{inconsolata}
\usepackage{booktabs}    
\usepackage{multirow}    
\usepackage{makecell}    

\newcommand\blfootnote[1]{%
  \begingroup
  \renewcommand\thefootnote{}\footnote{#1}%
  \addtocounter{footnote}{-1}%
  \endgroup
}
\usepackage{graphicx}

\newcommand{\modelname}{SegTune}

\title{SegTune: Structured and Fine-Grained Control for Song Generation}

\author{
  \textbf{Yuejiao Wang\textsuperscript{1*\ddag}},
  \textbf{Zihao Ji\textsuperscript{1*}},
  \textbf{Pengfei Cai\textsuperscript{2*\ddag}},
  \textbf{Xu Li\textsuperscript{1\dag}},
  \textbf{Haorui Zheng\textsuperscript{3\ddag}},
  \textbf{Zewen Song\textsuperscript{1}},
  \\
  \textbf{Zhongliang Liu\textsuperscript{1}},
  \textbf{Chen Zhang\textsuperscript{1}},
  \textbf{Pengfei Wan\textsuperscript{1}}
  \\
  \textsuperscript{1}Kling Team, Kuaishou Technology
  \\
  \textsuperscript{2}University of Science and Technology of China, Hefei, Anhui, China
  \\
  \textsuperscript{3}Peking University, Beijing, China
  \\
  \textsuperscript{*}Equal Contribution \quad
  \textsuperscript{\dag}Corresponding Author \quad
  \\yuejiaowang@link.cuhk.edu.hk, cqi525@mail.ustc.edu.cn, getsum@stu.pku.edu.cn
  \\ \{jizihao, lixu15, songzewen, dongliang06, zhangchen03, wanpengfei\}@kuaishou.com
}

\begin{document}
\maketitle
\blfootnote{$^{\ddag}$Work done during internship at Kling Team, Kuaishou Technology.}

\begin{abstract}
Recent advances in neural song generation have enabled high-quality synthesis from lyrics and global textual prompts. However, most systems fail to model temporally varying attributes of songs, severely limiting fine-grained control over musical structure and dynamics. To address this, we propose \textbf{\modelname{}}, a Diffusion Transformer-based framework enabling structured and fine-grained controllability by allowing users or large language models (LLMs) to specify local musical descriptions aligned to song segments. These segment prompts are temporally broadcast to corresponding time windows, while global prompts ensure stylistic coherence. To support precise lyric-to-music alignment, we introduce an LLM-based duration predictor that autoregressively generates sentence-level timestamps in LyRiCs format. We further construct a large-scale data pipeline for high-quality song collection with aligned lyrics and prompts, and propose new metrics to evaluate segment alignment and vocal consistency. Experiments demonstrate that \modelname{} outperforms existing baselines in both musicality and controllability. 
Visit our \href{https://github.com/KlingAIResearch/SegTune}{project page} for codes and more generated songs.
\end{abstract}

\section{Introduction}




Music and song constitute a powerful medium for emotional expression, combining linguistic content with rich acoustic and musical structures. Among music-related generative tasks, song generation is particularly challenging, as it requires the joint synthesis of vocals and accompaniment conditioned on lyrics and high-level control signals. Although commercial systems like Suno have demonstrated expert-level performance, the open-source community for song generation still has considerable room for technical advancement.

Early song generation systems predominantly adopt autoregressive (AR) transformers to model long-range dependencies over quantized audio tokens. Representative approaches such as SongCreator~\cite{songcreator} and MusiCoT~\cite{music_cot} employ a shared token vocabulary for vocals and accompaniment, but this design introduces modality interference and limits expressive capacity. Subsequent works—including YuE~\cite{yue}, SongGen~\cite{liu2025songgen}, and LeVo~\cite{lei2025levo}—mitigate this issue by modeling vocals and accompaniment as separate token sequences. Nevertheless, AR methods remain computationally expensive and inflexible for interactive editing. Alternatively, non-autoregressive (NAR) frameworks—including DiffRhythm~\cite{diffrhythm}, DiffRhythm$+$~\cite{diffrhythm_plus}, ACE-Step~\cite{acestepstep} and JAM~\cite{liu2025jamtinyflowbasedsong}—adopt diffusion or flow-matching for accelerated generation. By operating in latent audio spaces, these methods significantly reduce inference time while maintaining reasonable audio fidelity. However, NAR models face inherent challenges: they compress the full song generation pipeline (composition and rendering) into a single latent diffusion process. As a result, they often struggle to jointly optimize musical structure, temporal coherence and voice-instrument balance.

A fundamental limitation shared by both AR and NAR song generation systems is their predominant reliance on global-only control signals. This limitation manifests in three interrelated ways. First, global prompts fail to capture inherent temporal dynamics—attributes of music, such as instrumentation, emotion, and energy naturally evolve across song segments, leading to homogeneous and mediocre outputs. Although some approaches incorporate coarse structural tags into lyrics~\cite{yue, acestepstep, lei2025levo}, they still lack the resolution for truly fine-grained and segment-level controls. Second, under global-only conditioning, jointly generating vocals and accompaniment imposes a substantial coordination burden on the model, frequently leading to misaligned expression across modalities \cite{acestepstep}. Third, the absence of fine-grained control curtails expressive flexibility for both professional composers and amateur creators, which hampers practical usability for diverse creative workflows. These challenges are further exacerbated in some NAR models by their reliance on low-quality lyric durations—either zero-shot LLM-generated or manually specified by humans~\cite{diffrhythm, liu2025jamtinyflowbasedsong}. However, such duration annotations are not only time-consuming and error-prone, but also discourage user interaction.

To address these limitations, we propose \modelname{}, a NAR song generation framework that supports hierarchical control: a global prompt defines the overall style, while segment-level prompts—specifiable by users or auto-generated via an LLM—govern fine-grained per-segment attributes (e.g., emotion, rhythm, instrumentation). Specifically, we introduce a dedicated segment-level conditioning paradigm: a segment encoder injects fine-grained control signals into the corresponding temporal window of the latent sequence, while a global encoder preserves stylistic coherence across the entire song. Furthermore, we eliminate the need for manual lyric duration annotations by introducing a context-aware, LLM-based duration predictor, which adaptively generates sentence-level timestamps conditioned on the hierarchical prompts. 
Finally, these hierarchical textual conditionings and time-aligned lyrics embedding will guide the Diffusion Transformer (DiT) blocks to generate expressive and cohesively structured latent embeddings for both vocals and accompaniment.
In summary, our contributions are as follows:
\begin{itemize}[nosep]
\item 
We introduces a hierarchical, segment-level textual conditioning paradigm for fine-grained control in song generation.
\item 
We develop an LLM-based duration predictor that generates sentence-level lyric timestamps, enabling accurate lyrics alignment without manual annotation.
\item 
We construct a scalable pipeline to clean, annotate, and align high-quality songs with multi-level textual descriptions. 
\item 
We design new evaluation metrics, including segment-level MuLan score and singer attribute scoring, to rigorously assess fine-grained instruction following. 
\end{itemize}

\begin{figure*}[t]
    \centering
    \includegraphics[width=\linewidth]{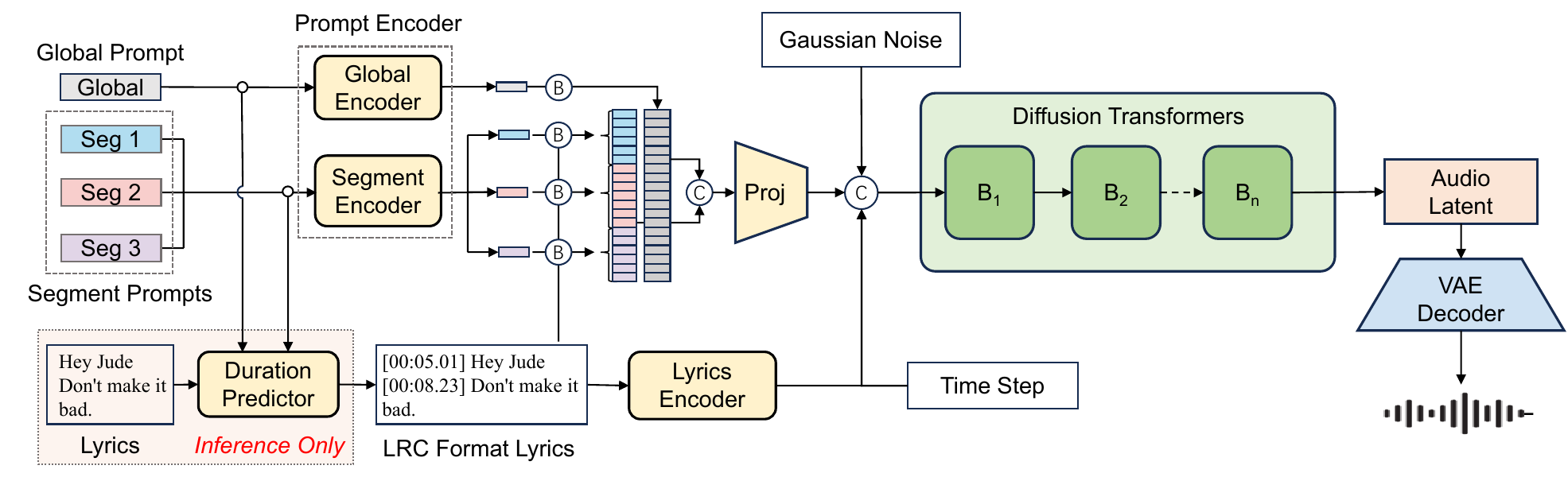}
    \caption{Overview of the \modelname{} architecture. The model takes lyrics and textual prompts as input. 
    An LLM-based duration predictor estimates sentence-level durations for the lyrics, while a lyrics encoder embeds the lyrics and performs sentence-level  alignment. 
    The prompt encoder encodes both global and segment prompts into 1D vectors. The global prompt is broadcast to all time steps, whereas each segment prompt is broadcast to the frames within its corresponding temporal window, as determined by the duration predictor.
    All the conditional embeddings are concatenated and fed into a Diffusion Transformer. In the diagram, \protect\circled{B} denotes temporal broadcasting within a segment window, and \protect\circled{C} denotes channel-wise feature concatenation.}
    
    \label{fig:main}
\end{figure*}

\section{Related Work}
\subsection{Music Generation}

Music generation focuses on producing coherent and stylistically consistent audio conditioned on text, melody, or other high-level cues. 
One prevalent approach is AR modeling, which sequentially generates discrete audio representations. 
MusicGen~\cite{musicgen} combines residual vector quantization with transformer-based decoder to improve fidelity and controllability, while MusicLM~\cite{musiclm} adopts a two-stage architecture that first predicts semantic tokens before rendering acoustics. 
MeLoDy~\cite{melody} enhances this design by replacing the acoustic decoder with a diffusion model to further improve synthesis quality and sampling efficiency.

In parallel, NAR methods~\cite{musicLDM, stableaudio, stableaudioopen} have emerged as promising alternatives. These models typically employ diffusion~\cite{DDPM, LDM} or flow-matching~\cite{flow-matching, I-CFM} mechanisms, operating in latent audio spaces to accelerate inference while maintaining high fidelity. 
Music ControlNet~\cite{music_control_net} extends this paradigm by introducing time-varying control signals—such as melody, rhythm, and dynamics—via temporally aligned conditioning, enabling fine-grained control over different aspects of the musical output. 
 TVC-MusicGen~\cite{yang2025tvc} adopts a  segment-level diffusion framework for fine-grained music generation, sharing similarities with our paradigm, but its application is limited to instrumental music without vocal track.

\subsection{Song Generation}
Song generation aims to synthesize full musical compositions, including vocals and accompaniment, based on input lyrics and optional control signals.
This task introduces additional challenges beyond instrumental  music generation, such as aligning melody with lyrical phrasing, keeping vocal–accompaniment coherence, and maintaining musically meaningful transitions across segments.

Early systems such as Jukebox~\cite{jukebox} adopt AR transformers to model long-range dependencies over quantized audio tokens. Later works, like SongCreator~\cite{songcreator} and MusiCoT~\cite{music_cot},  follow similar strategies but rely on a shared token vocabulary for both vocal and instrumental content, which introduces modality interference and limits expressive capacity. 
To mitigate this, more recent systems, including SongGen~\cite{liu2025songgen}, YuE~\cite{yue} and LeVo~\cite{lei2025levo}, model vocals and accompaniment as separate token sequences, enabling more faithful synthesis and better exploitation of language models for long-context generation. However, AR approaches suffer from slow inference, high training complexity, and limited flexibility in downstream tasks like song editing.

Alternatively, NAR frameworks, including DiffRhythm, DiffRhythm$+$, ACE-Step, and JAM~\cite{diffrhythm, diffrhythm_plus, acestepstep, liu2025jamtinyflowbasedsong}, leverage diffusion-based architectures for faster generation and easier extension. Nevertheless, these systems still face challenges to maintain musicality, long-range coherence, and balancing vocals with accompaniment, since both composition and acoustic rendering are handled within only DiT.

In particular, existing systems rely primarily on global text prompts, which are insufficient to capture the temporal variability of real songs.  
Some works~\cite{yue, acestepstep, lei2025levo} insert  structural segmental labels in lyrics to align with predefined musical segments, yet finer-grained local control, such as adjusting the instrumentation or emotional intensity of specific segments, remains unsupported.
In contrast, our work introduces fine-grained, segment-level textual conditioning along with an LLM-based duration prediction module. 
These designs enable precise control over the temporal evolution of musical attributes while supporting  textual prompts for each segment, paving the way for more expressive and controllable song generation.

\section{Methodology}

\subsection{Model Architecture}
As illustrated in Figure \ref{fig:main}, \modelname{} adopts a DiT~\cite{DIT} architecture based on conditional flow-matching, extending previous song generation models \cite{acestepstep,diffrhythm,diffrhythm_plus,liu2025jamtinyflowbasedsong}. The backbone consists of LLaMA-style~\cite{llama} transformer blocks. During training, a 1D VAE~\cite{stableaudio} is used to compress raw audio with a sampling rate of 44\,kHz into a latent sequence at 21.5\,Hz, which serves as the target trajectory for flow supervision. \modelname{} conditions generation on three complementary sources: global textual prompts, segment-level textual prompts, and time-aligned lyrics. Together, these signals control both the semantic content and the temporal evolution of musical attributes.

In the following subsections, we first detail the hierarchical conditioning mechanism, followed by the duration predictor, which generates precise lyric timestamps, overcoming the reliance on manual annotations in prior DiT-based approaches.

\subsubsection{Hierarchical prompts}
\modelname{} supports text prompts at both the global and segment levels. 
The global prompt controls high-level song attributes such as genre, gender, timbre, and the overall emotional tone. In contrast, segment-level prompts explicitly describe time-varying attributes, including segment structure tags (e.g., verse or chorus), emotional transitions, rhythmic patterns, and instrumentation. This separation allows the model to disentangle global stylistic consistency from local musical variation. Unlike prior approaches that encode structural information implicitly within lyrics or coarse tags, \modelname{} introduces explicit segment-level textual prompts that are independently encoded and temporally injected into the diffusion process. Examples of global and segment prompts can be found in our demo page.

Specifically, each segment prompt is encoded by a segment encoder into a vector $\mathbf{e}_s^{i} \in \mathbb{R}^{1 \times d_s}$, where $i$ is the $i$-th segment. This vector is temporally broadcast to all latent frames within the corresponding segment time window, ensuring consistent local conditioning. Noted that each segment time window was provided by the duration predictor module during the inference stage (refers to section \ref{sec:dur_predictor}).
Meanwhile, the global prompt is encoded by a global encoder and broadcast across all frames of the whole song.
The global and segment embeddings are concatenated along the channel dimension and projected through a three-layer MLP to obtain the final conditioning embedding $E_{\text{text}} \in \mathbb{R}^{T \times d_{\text{text}}}$, where $T$ is the length of the latent sequence and $d_{\text{text}} = 1024$. The detailed algorithm for text conditioning is shown in Appendix \ref{appe:algorithm}.

We employ Qwen3-Embedding-0.6B~\cite{qwen3embedding} as both the global and segment prompt encoder, as it preserves fine-grained semantic attributes in long-form textual descriptions.
Details of hierarchical prompt construction are provided in Section~\ref{pipeline}. During inference, segment-level prompts can be specified by users or automatically generated by a large language model, enabling flexible and expressive controls.

\subsubsection{Duration predictor}
\label{sec:dur_predictor}
Accurate duration predictor is a critical component of \modelname{}. It determines segment boundaries for prompt broadcasting, aligns lyrics with audio latents in temporal dimension, and defines the initial noise length during inference.

Despite its importance, duration prediction has been largely overlooked. Prior NAR works either require error-prone manual timestamps~\cite{acestepstep,diffrhythm,diffrhythm_plus} or use fragile zero-shot LLM prompting for word-level timing (e.g., JAM~\cite{liu2025jamtinyflowbasedsong}), neglecting the need for music-aware, lyric-aligned duration prediction module.

In this work, we fine-tune the Qwen3-4B-Base model~\cite{qwen3technicalreport}—selected for its favorable capacity–speed trade-off—as a duration predictor that generates sentence-level timestamps in LyRiCs (LRC) format. Specifically, in the inference stage, given lyrics together with global and segment prompts, the predictor outputs timestamped lyrics, learning to align durations with musical attributes such as rhythm, emotion, genre, and lyric length. The instruction template is detailed in Appendix \ref{appendix:duration prompt}, and the construction of ground-truth timestamps is described in Section~\ref{pipeline}. 
At inference time, segment temporal boundaries are derived directly from the predicted timestamps: for lyric-containing segments (e.g., verse and chorus), temporal boundaries are defined by the corresponding sentence intervals; for instrumental segments (e.g., intro, bridge and outro), boundaries are inferred from adjacent lyric segments, leveraging the structural contiguity inherent in typical song forms.

\subsubsection{Lyric conditioning}

To achieve precise, supervision-light phoneme-level alignment between lyrics and audio latent representations, we adopt the strategy of~\cite{diffrhythm,diffrhythm_plus} that only need sentence-start annotations.
Specifically, timestamped lyrics are directly available from the training data during training. In inference stage, the duration predictor automatically estimates these start times of lyrics. Each lyric sentence is first converted into a phoneme sequence via a grapheme-to-phoneme model \cite{wang2024maskgct}. Specifically, we used \textit{phonemizer} for English, and \textit{jieba} together with \textit{pypinyin} for Chinese. Then, a placeholder sequence $E_{\text{lyrics}}$, matching the length of the audio latent sequence $E_{\text{audio}}$, is initialized with \textit{<pad>} tokens. The phoneme sequence is written into $E_{\text{lyrics}}$ starting at the frame of the lyric’s start time. 

Finally, the textual prompt embedding $E_{\text{text}}$, along with $E_{\text{lyrics}}$, $E_{\text{audio}}$, and the time-step embedding $E_t$ (broadcast to length $T$) are concatenated along the channel dimension and fed into the DiT.

\subsection{Data Pipeline}
\label{pipeline}

\modelname{} is trained on a curated internal corpus of mainly Chinese pop songs and a small amount of other language songs. To ensure high-fidelity segment annotations and precise lyric-level timestamping, we devise a dedicated three-stage data curation pipeline, and the workflow is shown in the Appendix \ref{appendix:data pipeline}.

\subsubsection{Quality filtering}
We first apply metadata-based filtering (duration, sampling rate, channels, energy, etc.) and sound event detection to discard non-musical clips. Subsequently, we leverage Audiobox ~\cite{audiobox_Aesthetics} and SongEval~\cite{songeval} to score audio aesthetics and prune low-quality samples.

\subsubsection{Lyrics processing}
For songs without lyrics, we separate vocals using Demucs v4~\cite{demucs}, then transcribe them with FireRedASR~\cite{fireredasr} (Mandarin) or Whisper-Large-v3~\cite{whisper} (other languages). When ground-truth LRC files are available, we first remove non-lyrics metadata using an LLM-based filter, then validate the cleaned lyrics against ASR outputs via edit distance—discarding samples with high discrepancy. Then, structural labels (e.g., \textit{intro}, \textit{verse}, \textit{chorus}) are extracted using the all-in-one music understanding model~\cite{taejun2023allinone}.

\subsubsection{Hierarchical prompt annotation}
Global and segment prompts are generated via Audio Flamingo 3~\cite{audioflamingo3}, with the structural label prepended to each segment prompt to enable controllability. System prompt templates used for Audio Flamingo 3 are detailed in the Appendix \ref{appendix:AF3 prompt}. To mark boundaries, fixed prompts—“This piece is the start/end of the song.”—are assigned to the first and last 0.5\,s of each sample.

\subsection{Training and Inference}

The model is trained under the Conditional Flow Matching (CFM) framework, which aims to learn a function $v_\theta(t, C, x_t)$ that approximates the flow $u(x_t|x_0,x_1)$.  
The training objective is defined as:
\begin{gather}
\mathcal{L} = 
\mathbb{E}_{t,\, q,\, p} 
\left\| v_\theta(t, C, x_t) - u(x_t \mid x_0, x_1) \right\|^2,\\
x_t = (1 - t)x_0 + t x_1, \\
u(x_t \mid x_0, x_1) = x_1 - x_0,
\end{gather}
where $x_0 \sim p(x_0) $ represents a sample from the prior distribution $\mathcal{N}(0, \mathbf{I})$.
$x_1 \sim q(x_1)$ is drawn from the target data distribution,  
$t \sim \mathcal{U}(0, 1)$ denotes the diffusion time step,  
and $C$ denotes the conditioning input, which includes lyrics and textual prompts.  
The target vector $u(x_t \mid x_0, x_1)$ represents the flow at $x_t$.

Specifically, the training procedure follows a three-stage paradigm. (i) \textbf{Pre-training}:  
we retain songs with $\geq$32 kHz sampling rate, durations of 30s to 6 mins, and audio aesthetics scores above the 5th percentile, yielding approximately 370k songs (around 27k hours).
(ii) \textbf{Fine-tuning}: stricter criteria apply—44 kHz, stereo, and top-50\% audio aesthetics scores—resulting in about 50k songs (around 4k hours).
(iii) \textbf{Preference Alignment}: following prior work~\cite{lei2025levo,diffrhythm_plus,liu2025jamtinyflowbasedsong}, we adopt iterative Direct Preference Optimization (DPO)~\cite{dpo} with two rounds. Starting from the SFT model, each round generates 16 candidates per lyric; win–loss pairs are formed by selecting pairs with SongEval~\cite{songeval} score differences above a threshold, where the winning sample exceeds the 75th percentile of all candidates. Each DPO round uses around 20k such pairs.

During inference, we employ the Euler ODE solver.  
For Classifier-Free Guidance (CFG), we use the formulation:
\begin{align}
v = v_u + \text{cfg}(v_c - v_u) - \text{cfg}_n(v_n - v_u),
\end{align}
where $v_u$ and $v_c$ denote the unconditional and conditional flows, respectively,  
and $v_n$ represents the flow obtained under negative conditions~\cite{negative_prompt}.  
In the negative condition setup, the lyric conditioning is removed,  
while both global and local prompts are replaced with negative prompts.  
Empirically, we set $\text{cfg}=3$ and $\text{cfg}_n=1$.

\begin{table*}[t]  
\centering
\small
\caption{Performance comparison of SegTune and baseline models on objective metrics. SegTune-SFT denotes the model that has undergone both pretraining and supervised fine-tuning (SFT), while SegTune-DPO refers to the model further refined via 2 iterations of Direct Preference Optimization (DPO) starting from the SegTune-SFT checkpoint. G-Mulan denotes Global Mulan score, and S-Mulan denotes Segment Mulan score.}
\label{tab:performance_comparison}
\begin{tabular}{@{} 
    >{\centering\arraybackslash}p{1.9cm} 
    >{\centering\arraybackslash}p{0.7cm} 
    *{4}{>{\centering\arraybackslash}p{0.55cm}} 
    *{5}{>{\centering\arraybackslash}p{0.55cm}} 
    *{1}{>{\centering\arraybackslash}p{1.3cm}} 
    *{2}{>{\centering\arraybackslash}p{0.85cm}} 
    @{}}
\toprule
\multirow{3}{*}{\textbf{Models}}
& \multirow{3}{*}{\textbf{PER$\downarrow$}}
& \multicolumn{4}{c}{\textbf{AudioBox-aesthetic$\uparrow$}} 
& \multicolumn{5}{c}{\textbf{SongEval$\uparrow$}} 
& \multicolumn{3}{c}{\textbf{Instruction-following$\uparrow$}} \\
\cmidrule(lr){3-6} \cmidrule(lr){7-11} \cmidrule(lr){12-14}
& & \textbf{CE} & \textbf{CU} & \textbf{PC} & \textbf{PQ} 
& \textbf{Coh} & \textbf{Mem} & \textbf{NVBP} & \textbf{CSS} & \textbf{OM} 
& \textbf{G-Mulan} & \textbf{Gender} & \textbf{Age} \\
\midrule
YuE    & 48.5\% & 7.16 & 7.66 & 6.27 & 8.09 & 3.51 & 3.27 & 3.22 & 3.26 & 3.22 & 0.29 & 80.7\% & 44\% \\
LeVo    & 29.8\% & 7.43 & 7.71 & 5.25 & \underline{8.29} & 3.46 & 3.29 & 3.20 & 3.29 & 3.35 & 0.32 & \underline{90.6\%} & 50\% \\
DiffR.$+$ & 27.4\% & \underline{7.55} & \underline{7.80} & 6.72 & 8.21 & \underline{4.05} & \underline{3.84} & \underline{3.65} & \underline{3.82} & \underline{3.76} & \textbf{0.47} & 37.5\% & 54\% \\
ACE-Step & 35.6\% & 7.38 & 7.53 & 6.71 & 7.88 & 3.98 & 3.78 & \underline{3.65} & 3.77 & 3.74 & 0.35 & 78.1\% & \underline{56\%} \\
\midrule
SegTune-SFT & \textbf{14.5\%} & 7.38 & 7.71 & \textbf{6.83} & 8.23 & 3.54 & 3.22 & 3.23 & 3.32 & 3.19 & \textbf{0.47} & \textbf{96.7\%} & \textbf{57\%} \\
SegTune-DPO & \underline{18.5\%} & \textbf{7.63} & \textbf{7.85} & \underline{6.80} & \textbf{8.36} & \textbf{4.25} & \textbf{4.06} & \textbf{4.09} & \textbf{4.08} & \textbf{3.97} & \underline{0.46} & 81.0\% & 51\% \\
\bottomrule
\end{tabular}
\end{table*}

\section{Experiments}

\subsection{Experimental Setup}




\modelname{} is trained on an internal corpus of predominantly Mandarin pop songs ($>$90\%), spanning diverse artists, lyrical themes, and segment structure. The diffusion backbone is a DiT-style architecture (1.1B parameters, 16 LLaMA-style decoder blocks), following~\cite{diffrhythm}. Training proceeds in three stages: (i) 20-epoch pretraining with batch size = 32, lr = 2e-5, (ii) 8-epoch fine-tuning with the same setting, and (iii) two rounds of iterative DPO (4 epochs each, batch size = 8, grad accumulation = 4, and lr = 5e-7). During training, 20\% dropout is applied independently to global and segment-level conditions to enable classifier-free guidance. We also augment global and segment text prompts using LLM-based rewriting to enhance generalization across diverse real-world text input styles. 

For the duration predictor, we fine-tune Qwen3-4B-Base on $>$100k LRC-formatted lyrics for 8 epochs (batch size = 8, grad accumulation = 4, max new tokens = 4096, and lr = 2e-5), using LoRA~\cite{hu2022lora} (rank = 32) for efficiency.

\subsection{Baselines and Evaluation}

We select four representative state-of-the-art baselines: YuE~\cite{yue} and LeVo~\cite{lei2025levo}, which adopt AR language models; DiffRhythm$+$~\cite{diffrhythm_plus} and ACE-Step~\cite{acestepstep}, which are diffusion-based models. All baselines support song generation conditioned on lyrics and global textual prompts.
And the test set for song generation consists of 15 Mandarin pop songs generated by ChatGPT, including their lyrics and associated prompts. To ensure strict robustness, we generated 10 unique audio samples per prompt for every model, totaling 150 generated tracks per system. And we provided standard deviations for objective metrics to demonstrate the stability of different systems.

We evaluate aesthetic quality using objective metrics: (i) Phoneme Error Rate (PER), evaluating the intelligibility and lyrical fidelity of songs. FireRedASR \cite{fireredasr} was used to transcribe generated songs; (ii) Audiobox-Aesthetics~\cite{audiobox_Aesthetics}, assessing production quality (PQ), production complexity (PC), content enjoyment (CE), and content usefulness (CU); (iii) SongEval~\cite{songeval}, measuring coherence (Coh), memorability (Mem), natural vocal breathing/phrasing (NVBP), clarity of song structure (CSS), and overall musicality (OM).

The instruction-following capability of music generation models is then evaluated using the Muq-MuLan~\cite{zhu2025muq} score. Muq-Mulan is a widely adopted multimodal representation model that aligns text descriptions and music clips through contrastive learning, and has been extensively used for music evaluation \cite{acestepstep,yue}. Global Mulan score measures the global alignment between songs and their global prompts. In addition, we compute the Segment MuLan scores for individual song segments based on segment prompts and use their average to assess the model’s segment instruction adherence.
We further evaluate the model’s instruction-following accuracy on singer-related attributes—gender and age—since MuLan shows limited sensitivity to vocal identity.
Prompts are selected from the test set, and the singer’s gender and age (teenager, 20s, and 40s) are modified to form new global prompts. For gender accuracy, we directly employ Qwen3-Omni-30B-A3B-Captioner~\cite{Qwen3-Omni} to assess the gender of generated songs. For age accuracy, we randomly sample two generated songs and conduct an A/B test, where Qwen3-Omni determines which song corresponds to the older singer. The final accuracy is computed based on the correctness of these pairwise judgments.

For subjective evaluation, we conducted 5-scale Mean Opinion Score (MOS) listening tests, in which five listeners rated each sample on musicality and quality using fine-grained evaluation criteria. To ensure balanced evaluation, every listener assessed all 5 system outputs for each of 9 songs—totaling 45 ratings per listener. This scale aligns closely with recent foundational music models (which typically use 15–20 prompts due to human cognitive load). The sample size is fully sufficient to prove statistical significance.

\section{Results and Discussions}

\subsection{Results of Objective Evaluation}

Table~\ref{tab:performance_comparison} presents objective results for SegTune (SFT and DPO stages) and baselines. SegTune achieves strong performance overall: it has the lowest PER, indicating superior lyrical fidelity and vocal intelligibility. In contrast, we noticed that the vocal generated by YuE is a bit hoarse. Both YuE and ACE-Step frequently interpolate past lyrics into current segments, resulting in notably higher PER. As for AudioBox-Aesthetics and SongEval metrics, SegTune-SFT is competitive with baselines, and SegTune-DPO further enhances quality—particularly in SongEval metrics. More detailed results for standard deviations are reported in Appendix \ref{appendix:std}.

Since baseline models support only global control, we assess instruction-following via global MuLan scores and singer attribute control accuracy (gender and age). SegTune-SFT attains the highest gender control accuracy while preserving strong global MuLan alignment, demonstrating superior instruction-following capability. 

After DPO fine-tuning, we observe a trade-off: the musical quality improves, yet the instruction-following degrades, especially for age and gender. This stems from bias in the preference data and DPO constrains: winning songs scored by SongEval are dominated from young female vocals, and DPO optimizes only for preference alignment without enforcing age and gender fidelity. The limitation of DPO doesn't affect the core contribution of Segtune itself, which solves the long-standing lack of fine-grained control in NAR models. One promising direction is to adopt online policy optimization, wherein generated songs are dynamically evaluated for demographic fidelity, and deviations from user-specified attributes are explicitly penalized, which will be investigated in the future work.

\subsection{Results of Subjective Evaluation}
\begin{figure}[!t]
    \centering
    \includegraphics[width=\linewidth]{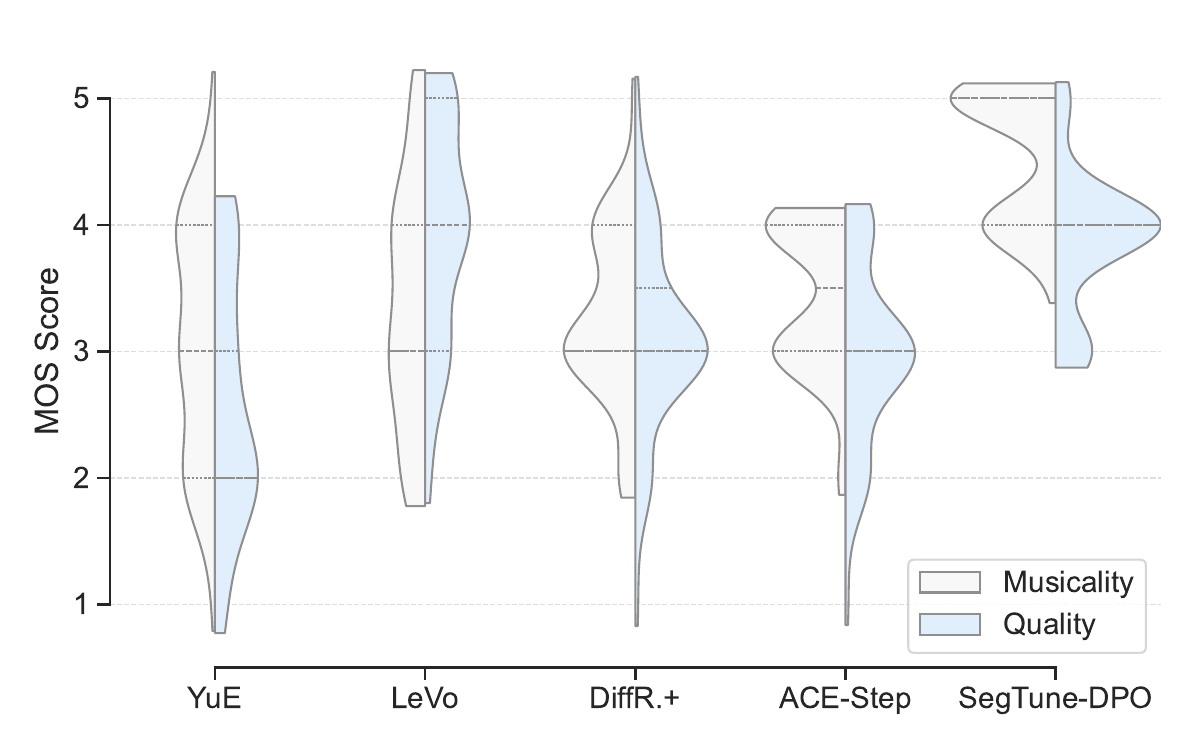}
    
    \caption{Violin plots of MOS results for musicality and quality evaluation. }
    
    \label{fig:mos}
\end{figure}

As shown in Figure ~\ref{fig:mos}, SegTune-DPO achieves the highest MOS for musicality among all baselines, with the lowest standard deviation ($4.57\pm0.52$)—indicating not only superior musical expressiveness but also exceptional consistency across samples. This gain is attributable to the DPO fine-tuning stage, which effectively suppresses musically degraded outputs. 
In contrast, YuE and LeVo exhibit higher variability in Musicality scores (standard deviations $>$ 0.9), as they heavily rely on audio prompts for conditioning, while the use of text-based global prompts leads to a marked degradation in accompaniment coherence and diversity.

As for quality, SegTune-DPO achieves the second-highest results, while exhibiting smallest standard deviation ($3.87\pm0.56$), surpassed only by LeVo ($3.96\pm0.87$). This is likely attributable to our data pipeline, which selects high quality training data, and to the post-training stages, which effectively enhances audio fidelity. As shown in the Wilcoxon signed-rank test result in Appendix \ref{appendix:Wilcoxon}, the superiority of SegTune-DPO over baselines in Musicality and Quality is highly significant (p<.001).

\subsection{Ablation Studies for Prompt Encoder}
\label{sec:ablation_encoder}

Since SegTune differs from baselines in data, scale, and architecture, the gains in Table~\ref{tab:performance_comparison} may reflect confounding factors. To isolate the impact of segment prompt injection, we perform ablation studies with identical data, backbone, and hyperparameters—varying only the prompt encoder.
 
Specifically, we examine two design dimensions:
(i) composition strategies for integrating embeddings from the hierarchical prompt encoders—namely, global-only, concatenate, or mixed settings;
and (ii) the choice of prompt encoder backbone—comparing Qwen3-Embedding, a state-of-the-art textual encoder optimized for long-form, natural-language prompts, against Muq-MuLan~\cite{zhu2025muq}, the music-specialized multimodal encoder widely adopted in recent song generation systems~\cite{diffrhythm_plus, liu2025jamtinyflowbasedsong, yang2025tvc}.

\begin{table*}[t]
\centering
\small
\caption{Impact of prompt encoder settings on objective performance of SegTune at the SFT stage. MuQ refers to MuQ-Mulan encoder, and Qwen3. refers to Qwen3-Embedding. G-Mulan denotes Global Mulan score, and S-Mulan denotes Segment Mulan score.}
\label{tab:performance_comparison_ablation}

\begin{tabular}{@{}
    >{\centering\arraybackslash}p{0.85cm}
    >{\centering\arraybackslash}p{0.85cm}
    *{4}{>{\centering\arraybackslash}p{0.4cm}}
    *{5}{>{\centering\arraybackslash}p{0.5cm}}
    *{1}{>{\centering\arraybackslash}p{1.25cm}}
    *{1}{>{\centering\arraybackslash}p{1.2cm}}
    *{2}{>{\centering\arraybackslash}p{0.75cm}}
    @{}}
\toprule
\multicolumn{2}{l}{\textbf{Prompt encoders}} 

& \multicolumn{4}{c}{\textbf{AudioBox-aesthetic$\uparrow$}} 
& \multicolumn{5}{c}{\textbf{SongEval$\uparrow$}} 
& \multicolumn{4}{c}{\textbf{Instruction-following$\uparrow$}} \\
\cmidrule(lr){1-2} \cmidrule(lr){3-6} \cmidrule(lr){7-11} \cmidrule(lr){12-15}

\textbf{Global} & \textbf{Segment} 
& \textbf{CE} & \textbf{CU} & \textbf{PC} & \textbf{PQ} 
& \textbf{Coh} & \textbf{Mem} & \textbf{NVBP} & \textbf{CSS} & \textbf{OM} 
& \textbf{G-Mulan} & \textbf{S-Mulan} & \textbf{Gender} & \textbf{Age} \\
\midrule

MuQ & --     & \underline{7.42} & 7.60 & 6.63 & 8.19 & 3.18 & 2.87 & 2.81 & 2.93 & 2.86 & 0.39 & 0.30 & 47.6\% & 47\% \\
Qwen3. & --  & 7.39 & 7.64 & \underline{6.76} & 8.19 & 3.42 & 3.11 & 3.11 & 3.20 & 3.12 & 0.40 & 0.33 & \underline{92.2\%} & 50\% \\
\midrule
\textbf{Concat.} \\
Qwen3. & MuQ & \textbf{7.57} & \textbf{7.82} & 6.63 & \textbf{8.35} & \textbf{3.62} & \textbf{3.37} & \textbf{3.30} & \textbf{3.43} & \textbf{3.34} & \underline{0.44} & \underline{0.37} & 84.4\% & 46\% \\
Qwen3. & Qwen3. & 7.38 & 7.71 & \textbf{6.83} & 8.23 & \underline{3.54} & \underline{3.22} & \underline{3.23} & \underline{3.32} & \underline{3.19} & \textbf{0.47} & \textbf{0.38} & \textbf{96.7\%} & \textbf{57\%} \\
\midrule
\textbf{Mixed} \\
Qwen3. & Qwen3. & 7.29 & \underline{7.73} & 6.33 & \underline{8.32} & 3.43 & 3.14 & 3.15 & 3.23 & 3.12 & 0.43 & 0.35 & 90.5\% & \textbf{57\%} \\
\bottomrule
\end{tabular}
\end{table*}

\begin{table*}[t]
\centering
\small
\vspace{1em}
\caption{Impact of duration predictor on SegTune-DPO performance. MAE denotes the mean absolute error (in seconds) of sentence-level duration prediction. GT means using the ground truth timestamps of lyrics for model inference. G-Mulan denotes Global Mulan score, and S-Mulan denotes Segment Mulan score.}
\label{tab:dur_predictor}
\begin{tabular}{@{}
    >{\centering\arraybackslash}p{1.6cm}
    >{\centering\arraybackslash}p{0.65cm}
    *{4}{>{\centering\arraybackslash}p{0.4cm}}
    *{5}{>{\centering\arraybackslash}p{0.5cm}}
    *{1}{>{\centering\arraybackslash}p{1.25cm}}
    *{1}{>{\centering\arraybackslash}p{1.2cm}}
    *{2}{>{\centering\arraybackslash}p{0.7cm}}
    @{}}
\toprule
\addlinespace[4pt]
\multirow{3}{*}{\textbf{Predictor}}
& \multirow{3}{*}{\textbf{MAE$\downarrow$}}
& \multicolumn{4}{c}{\textbf{AudioBox-aesthetic$\uparrow$}} 
& \multicolumn{5}{c}{\textbf{SongEval$\uparrow$}} 
& \multicolumn{4}{c}{\textbf{Instruction-following$\uparrow$}} \\
\cmidrule(lr){3-6} \cmidrule(lr){7-11} \cmidrule(lr){12-15}
& & \textbf{CE} & \textbf{CU} & \textbf{PC} & \textbf{PQ} 
  & \textbf{Coh} & \textbf{Mem} & \textbf{NVBP} & \textbf{CSS} & \textbf{OM} 
  & \textbf{G-Mulan} & \textbf{S-Mulan} & \textbf{Gender} & \textbf{Age} \\
\midrule
GT        & 0.00 & 7.65 & 7.74 & 6.58 & 8.35 & \textbf{4.33} & 4.16 & 4.17 & 4.12 & 4.01 & 0.45 & 0.47 & 81.9\% & 61\% \\
Qwen3-SFT & 0.99 & 7.66 & 7.74 & \textbf{6.58} & \textbf{8.36} & 4.32 & \textbf{4.16} & \textbf{4.18} & \textbf{4.16} & \textbf{4.06} & 0.45 & 0.41 & 81.9\% & 61\% \\
GPT-4o    & 3.24 & \textbf{7.69} & \textbf{7.76} & 6.29 & 8.34 & 4.19 & 4.00 & 4.08 & 3.98 & 3.86 & 0.42 & 0.41 & 81.9\% & 58\% \\
\bottomrule
\end{tabular}

\end{table*}

\subsubsection{Global-only setting} 
We train variants of SegTune-SFT using only global prompts with either Qwen3-Embedding or Muq-MuLan~\cite{zhu2025muq} as the global encoder. As Table~\ref{tab:performance_comparison_ablation} shows, Qwen3-Embedding consistently surpasses Muq-MuLan in both musicality and instruction-following. Notably, gender control accuracy reaches 92.2\% with Qwen3-Embedding, versus near-chance performance with Muq-MuLan. This discrepancy stems primarily from the fact that during MuQ-Mulan's text–music alignment training, singer-related attributes were not included in music captions, resulting in a poor representation of vocal characteristics in its embeddings. We conduct controlled experiments and visualization analyses on singer gender control to further elucidate the limitations of MuQ-MuLan in Appendix \ref{appendix:gender control}.

\vspace{-0.6em}
\subsubsection{Concatenate setting} In this setting, global and segment embeddings are concatenated along the channel dimension. Since Qwen3-Embedding achieves high-fidelity control over singer attributes, we fix it as global encoder and compare two choices for segment encoder: Muq-MuLan~\cite{zhu2025muq} and Qwen3-Embedding. Although Muq-MuLan encodes singer-related characteristics weakly, it remains suitable for segment-level control—where prompts primarily govern musical attributes such as instrumentation, emotion, and rhythm. 
In Table~\ref{tab:performance_comparison_ablation}: (i) the two variants achieve comparable overall musicality, yet the configuration using Qwen3-Embedding for the segment encoder yields stronger instruction-following performance; (ii) all objective musicality metrics significantly surpass those of the global-only setting. These results validate the superiority of hierarchical prompts in SegTune.

\subsubsection{Mixed setting} 
Following~\cite{jiang2025freeaudio}, we also explore linearly fuse global and segment embeddings (with weights 0.2 and 0.8), but observe clear drops in musicality and instruction following—likely due to semantic interference from mixing, unlike the cleaner separation in concatenation.

\subsection{Ablation Studies for Duration Predictor}
\label{sec:ablation_predictor}
An accurate duration predictor yields more musically plausible lyric timestamps than user-provided ones, lowering user expertise barriers. We evaluate two approaches: (i) a Qwen3-4B-Base model fine-tuned on our dataset to predict per-line start timestamps; and (ii) a zero-shot GPT-4o predictor with engineered prompting, following JAM~\cite{liu2025jamtinyflowbasedsong}.  
This evaluation uses 15 real-world Mandarin pop songs with paired lyrics and prompts, to enable direct comparison with ground-truth timestamps. To ensure fairness, these songs are excluded from the training set.

As shown in Table~\ref{tab:dur_predictor}, the Qwen3-SFT predictor achieves a mean absolute error (MAE) of 0.99s, significantly lower than GPT-4o’s. It also matches or exceeds GPT-4o across nearly all musicality dimensions, with comparable scores only in Content Enjoyment and Usefulness.
Segment MuLan and gender control remain largely unaffected, indicating robustness to local timing variations.


\section{Conclusions}

We presented \modelname{}, a NAR framework for controllable song generation. By jointly leveraging segment-level textual conditioning and LLM-based duration predictor, it enables fine-grained control over emotion, rhythm, and instrumentation while preserving global coherence. Coupled with a scalable data pipeline and comprehensive evaluation, \modelname{} significantly advances controllability and musical quality over existing open-source systems. Although our experiments focus on Mandarin pop songs, \modelname{} is language-agnostic and can be extended to other languages given appropriate data.

In the future work, we will explore: (i) supporting richer local interactions—e.g., multi-singer transitions—currently limited by data scarcity; and (ii) integrating \modelname{} into a conversational agent, which can interpret user intent and dynamically generate lyrics and hierarchical prompts for human-machine collaborative creation.

\section{Limitations}
While \modelname{} advances fine-grained controllability in non-autoregressive song generation, several limitations remain.

First, \modelname{} is trained on ground-truth songs with clearly defined segment structures. Consequently, when the segment structure of the user input is ambiguous or poorly specified, the performance of the duration predictor degrades, which will in turn negatively affect the overall song generation quality. To mitigate this, commercial applications typically integrate structured input templates and a prompt enhancer (PE) module that reformulates free-form user inputs into the consistent formats used during training. While this work prioritizes the core model architecture over interface engineering, we recognize that robust preprocessing pipelines are essential for bridging research prototypes with reliable, user-facing deployment.

Second, while the hierarchical prompting scheme allows control over segment-level attributes such as emotion and instrumentation, it cannot model fine-grained intra-segment dynamics (e.g., gradual crescendos or vocal ornamentation). 

We view these limitations not as fundamental flaws of SegTune, but as clear pathways for future work—such as integrating end-to-end structure prediction and developing multi-objective alignment strategies that preserve both musicality and user-specified attributes.

\section{Social Impact}

The democratization of song generation significantly lowers the barrier to music creation, enabling broader public participation in audio production. However, this accessibility may also intensify market competition for professional musicians and composers. We envision a collaborative paradigm where artists integrate AI generation tools into existing workflows to augment productivity, accelerate prototyping, and expand their expressive capabilities, rather than being displaced by them.

Concurrently, the proliferation of generative music models necessitates careful attention to copyright and vocal likeness rights. To mitigate unauthorized vocal replication, deployment platforms should restrict the use of professional singers' vocal samples as reference inputs and establish clear usage guidelines. Addressing melodic similarity requires a multi-stage approach: (1) curating training datasets to exclude copyrighted compositions, and (2) deploying post-generation similarity screening mechanisms to detect and filter outputs that closely resemble protected melodic material. We advocate for ongoing collaboration among researchers, industry practitioners, and policymakers to develop transparent governance frameworks that balance creative innovation with the protection of creators' intellectual property.

\section*{Acknowledgments}
We thank the following colleagues for their support and contributions to this work (sorted by alphabetical order): Youjun Chen, Qianyue Hu, Teng Ma, Junjie Yan.

\bibliography{custom}

\appendix


\section{Algorithm for Global and Segment Text Conditioning}
\label{appe:algorithm}

The algorithm \ref{alg_train} and algorithm \ref{alg_inference} for extracting global and segment-level text conditioning differs slightly between training and inference stages. 
During training, the start time of each lyric line and the temporal boundaries of each segment are pre-annotated by the data pipeline and thus directly available for model training. In contrast, during inference, the input lyrics provided by the user to SegTune contain no timestamps. Consequently, the duration predictor module is employed to estimate the start times of individual lyric lines and the time spans of each segment.

\begin{algorithm}[tp]
\caption{Global and Segment Text Conditioning Embedding Extraction during Training Stage}
\label{alg_train}

\begin{algorithmic}
\REQUIRE Global textual prompt $x_g$, segment textual prompts and the corresponding time boundary $\{(x_s^{i}, t_{s}^{i}, t_e^{i})\}_{i=1}^N$, where $i$ represent the $i$-th segment, $t_s$ and $t_e$ denote the start and end time of the $i$-th segment.
\REQUIRE Sampling rate $r$, downsample rate $r_d$, number of latent frames $T$, global encoder $f_g$, segment encoder $f_s$, output projection $\texttt{out\_proj}$

\STATE $\mathbf{e}_g \leftarrow f_g(x_g) \in \mathbb{R}^{1 \times d_g}$
\STATE $E_g \leftarrow \texttt{repeat}(\mathbf{e}_g, T)  \in \mathbb{R}^{T \times d_g} $   \hfill // Broadcast the global prompt across all time frames.  

\STATE Initialize $E_s \in \mathbb{R}^{T \times d_s}$ with zeros

\FOR{each $ (x_s^{i},t_{s}^{i}, t_e^{i})$ in segment prompts}
    \STATE $j_s^{i} \leftarrow \lfloor t_s^{i} \cdot r / r_d \rfloor$, $i_e^{i} \leftarrow \lfloor t_e^{i} \cdot r / r_d \rfloor$
    \STATE $\mathbf{e}_s^{i} \leftarrow f_s(x_s^{i}) \in \mathbb{R}^{1 \times d_s}$
    \STATE $E_s[j_s^{i} : j_e^{i}] \leftarrow \mathbf{e}_s^{i}$   // Broadcast each segment prompt to its corresponding temporal window. 
\ENDFOR

\STATE $E_{\text{cat}} \leftarrow \texttt{concat}(E_g, E_s, \text{dim}=-1)   $, $E_{\text{cat}} \in \mathbb{R}^{T \times (d_g + d_s)}$
\STATE $E_{text} \leftarrow \texttt{out\_proj}(E_{\text{cat}})$
\RETURN fused embedding $E_{text} \in \mathbb{R}^{T \times d_{text}}$
\end{algorithmic}
\end{algorithm}

\begin{algorithm}[tp]
\caption{Global and Segment Text Conditioning Embedding Extraction during Inference Stage}
\label{alg_inference}

\begin{algorithmic}
\REQUIRE Global textual prompt $x_g$, segment textual prompts $\{x_s^{i}\}_{i=1}^N$, where $i$ represent the $i$-th segment.
\REQUIRE Sampling rate $r$, downsample rate $r_d$, global encoder $f_g$, segment encoder $f_s$, duration predictor module $f_p$, output projection $\texttt{out\_proj}$

\STATE $\{(t_{s}^{i}, t_e^{i})\}_{i=1}^N, T \leftarrow f_p(\{x_s^{i}\}_{i=1}^N, x_g)$
\STATE $\mathbf{e}_g \leftarrow f_g(x_g) \in \mathbb{R}^{1 \times d_g}$
\STATE $E_g \leftarrow \texttt{repeat}(\mathbf{e}_g, T)  \in \mathbb{R}^{T \times d_g} $   \hfill // Broadcast the global prompt across all time frames.  

\STATE Initialize $E_s \in \mathbb{R}^{T \times d_s}$ with zeros

\FOR{each $ (x_s^{i},t_{s}^{i}, t_e^{i})$ in segment prompts}
    \STATE $j_s^{i} \leftarrow \lfloor t_s^{i} \cdot r / r_d \rfloor$, $i_e^{i} \leftarrow \lfloor t_e^{i} \cdot r / r_d \rfloor$
    \STATE $\mathbf{e}_s^{i} \leftarrow f_s(x_s^{i}) \in \mathbb{R}^{1 \times d_s}$
    \STATE $E_s[j_s^{i} : j_e^{i}] \leftarrow \mathbf{e}_s^{i}$   // Broadcast each segment prompt to its corresponding temporal window. 
\ENDFOR

\STATE $E_{\text{cat}} \leftarrow \texttt{concat}(E_g, E_s, \text{dim}=-1)   $, $E_{\text{cat}} \in \mathbb{R}^{T \times (d_g + d_s)}$
\STATE $E_{text} \leftarrow \texttt{out\_proj}(E_{\text{cat}})$
\RETURN fused embedding $E_{text} \in \mathbb{R}^{T \times d_{text}}$
\end{algorithmic}
\end{algorithm}

\section{Prompt of Duration Predictor} \label{appendix:duration prompt}


\begin{tcolorbox}[
  enhanced,
  colback=gray!5,
  colframe=black,
  coltitle=white,
  colbacktitle=black,
  title=\textbf{Input Prompt to Qwen3-4B-Base},
  breakable,
  sharp corners,
  boxrule=0.5pt
]
You are a professional music composer and vocal arranger.
Your task:

1. Analyze the lyrics and the song description below.

2. For each line of lyrics, estimate a reasonable singing duration. Base your estimation jointly on:
\begin{itemize}
    \item The intrinsic characteristics of the line itself (e.g., length, phrasing, complexity)
    \item The overall song attributes;
    \item The structural flow of the song, including instrumental breaks, natural pauses, and transitions;
\end{itemize}

3. Return: Output a complete `.lrc` style list with timestamps.     \\

Below are the target global song description and lyrics. Please follow the instructions above and return the completed .lrc file directly.
\\

\textbf{Song Description}

This pop rock ballad features a male vocalist delivering an emotional and uplifting melody. The mood is warm and introspective, with a gradually intensifying energy that enhances the song’s heartfelt tone. The singer’s voice is soulful and expressive, using subtle dynamic shifts to convey a sense of comfort and encouragement. 
\\

\textbf{Lyrics}

[This piece is the start of the song.]\\

[This piece is the intro of the song. The song segment features a spare and gentle piano motif, setting a contemplative and soothing mood ...
]\\

[This piece is the first verse of the song. The segment features a tender vocal performance accompanied by a steady, melodic piano line ...
]

Hey Jude, don't make it bad, \\
Take a sad song and make it better.\\
...\\

[This piece is the second verse of the song. The segment features a tender vocal performance accompanied by a steady, melodic piano line ...
]

Hey Jude, don't be afraid, \\
You were made to go out and get her.\\
...
\\

LRC Prediction:

\end{tcolorbox}

\section{Workflow of Data Pipeline}
\label{appendix:data pipeline}

\begin{table*}[!h]
  \centering
  \caption{Objective metrics averaged across 150 generated tracks per system, with mean $\pm$ standard deviation.}
  \label{tab:objective_results}
  \small
  \setlength{\tabcolsep}{2.8pt} 
  \begin{tabular}{@{}l *{9}{c}@{}}
    \toprule
    \textbf{Model} & \textbf{CE}$\uparrow$ & \textbf{CU}$\uparrow$ & \textbf{PC}$\uparrow$ & \textbf{PQ}$\uparrow$ & \textbf{Coh}$\uparrow$ & \textbf{Mem}$\uparrow$ & \textbf{NVBP}$\uparrow$ & \textbf{CSS}$\uparrow$ & \textbf{OM}$\uparrow$ \\
    \midrule
    YuE         & 7.16$\pm$.60 & 7.66$\pm$.23 & 6.27$\pm$1.54 & 8.09$\pm$.14 & 3.51$\pm$.35 & 3.27$\pm$.38 & 3.22$\pm$.34 & 3.26$\pm$.38 & 3.22$\pm$.36 \\
    LeVo        & 7.43$\pm$.55 & 7.71$\pm$.15 & 5.25$\pm$1.83 & 8.29$\pm$.10 & 3.46$\pm$.39 & 3.29$\pm$.40 & 3.20$\pm$.32 & 3.29$\pm$.36 & 3.35$\pm$.34 \\
    DiffR.+     & 7.55$\pm$.05 & 7.80$\pm$.04 & 6.72$\pm$.06 & 8.21$\pm$.08 & 4.05$\pm$.08 & 3.84$\pm$.12 & 3.65$\pm$.10 & 3.82$\pm$.11 & 3.76$\pm$.12 \\
    ACE-Step    & 7.38$\pm$.08 & 7.53$\pm$.03 & 6.71$\pm$.03 & 7.88$\pm$.09 & 3.98$\pm$.04 & 3.78$\pm$.05 & 3.65$\pm$.06 & 3.77$\pm$.05 & 3.74$\pm$.05 \\
    SegTune-SFT & 7.38$\pm$.03 & 7.71$\pm$.03 & \textbf{6.83}$\pm$.05 & 8.23$\pm$.02 & 3.54$\pm$.05 & 3.22$\pm$.07 & 3.23$\pm$.07 & 3.32$\pm$.06 & 3.19$\pm$.06 \\
    SegTune-DPO & \textbf{7.63}$\pm$.02 & \textbf{7.85}$\pm$.02 & 6.80$\pm$.04 & \textbf{8.36}$\pm$.01 & \textbf{4.25}$\pm$.02 & \textbf{4.06}$\pm$.03 & \textbf{4.09}$\pm$.03 & \textbf{4.08}$\pm$.03 & \textbf{3.97}$\pm$.03 \\
    \bottomrule
  \end{tabular}
\end{table*}

\begin{figure}[h]
    \centering
    \includegraphics[width=1\linewidth]{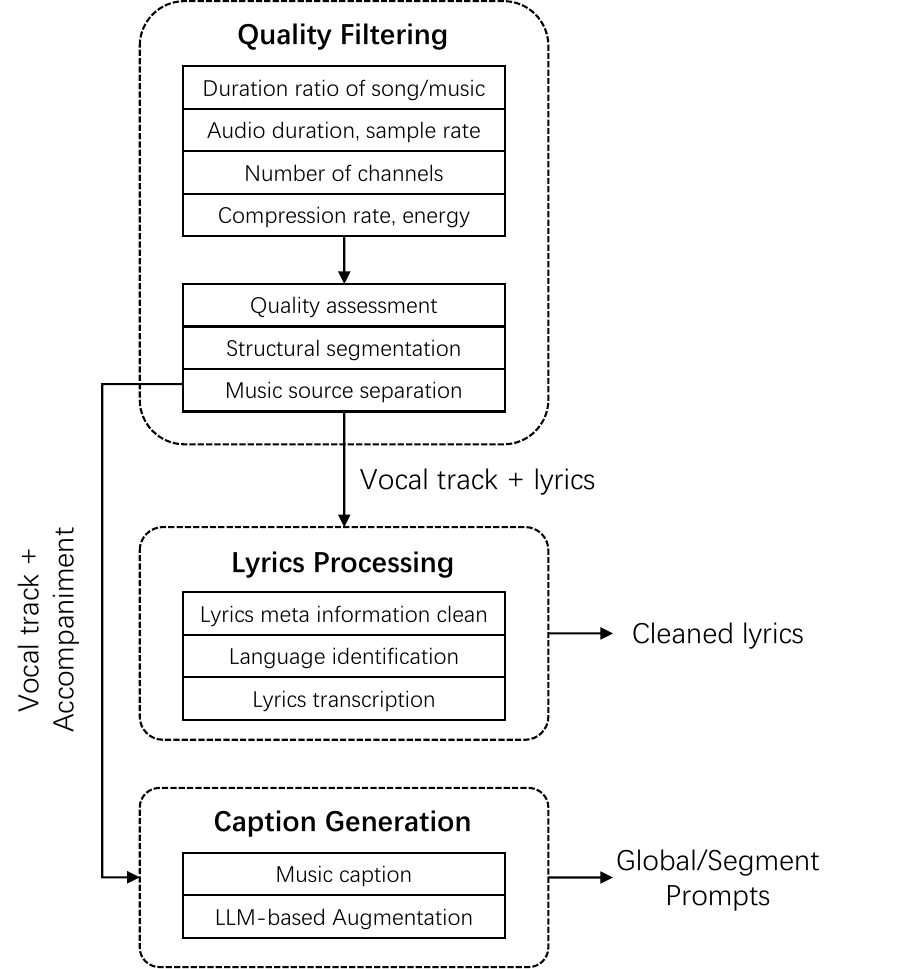}
    \caption{Overview of the data pipeline of \modelname{}.}
    \label{fig:data-pipeline}
\end{figure}

Figure \ref{fig:data-pipeline} shows the data curation pipeline, which comprises three key stages: 
(1) Quality filtering, where non-musical or low-quality clips are removed via metadata constraints and aesthetic scoring (using Audiobox Aesthetics and SongEval); (2) Lyrics processing, involving multilingual ASR transcription, and alignment validation against ground-truth LRC files via edit distance; and (3) Hierarchical music caption generation, where global and segment-level prompts are generated by Audio Flamingo 3, augmented with structural labels for controllability, and boundary markers are inserted at the start/end of each sample.

\section{Audio-Flamingo 3 Caption Prompt} 
\label{appendix:AF3 prompt}

\begin{tcolorbox}[
  enhanced,
  colback=gray!5,
  colframe=black,
  coltitle=white,
  colbacktitle=black,
  title=\textbf{Prompt for Global Caption Generation},
  breakable,
  sharp corners,
  boxrule=0.5pt
]

You are a helpful AI assistant. You need to act as a caption generator for music and generate descripitons in MusicCaps style. Describe the music in vivid detail, using the following rules: 
\begin{enumerate}
    \item Describe the details about genre, mood, feeling, ambience, and other notable features of the music.
    \item Describe the singer's  vocal characteristics, including gender, age range, vocal timbre, pitch range, and other notable features of the singer.
    \item Keep the descripiton within 1-4 sentences.
    \item Only provide details you are confident about. It is not compulsory to provide all details, but do not hallucinate.
\end{enumerate}
\end{tcolorbox}


\begin{tcolorbox}[
  enhanced,
  colback=gray!5,
  colframe=black,
  coltitle=white,
  colbacktitle=black,
  title=\textbf{Prompt for Segment Caption Generation},
  breakable,
  sharp corners,
  boxrule=0.5pt
]

You are a helpful AI assistant. Describe the song segment as part of a complete piece of song in vivid detail according to what you hear. Generate the descripiton using the following rules: 
\begin{enumerate}
    \item Include the instrumentation, rhythm and melody style, mood, emotional's impact, intensity and change.
    \item Mention any notable singing and playing techniques that occur and dynamic changes of the song.
    \item Keep the descripiton within 1-3 sentences.
    \item Only provide details you are confident about. It is not compulsory to provide all details, but do not hallucinate.
    
\end{enumerate}

\end{tcolorbox}





\newpage
\section{Objective Metrics of Music Generation Systems with Standard Deviation } 
\label{appendix:std}
To ensure rigorous evaluation, we generated 10 unique audio samples per prompt for each model, yielding 150 tracks per system for metric averaging. Standard deviations are now reported for objective metrics (the same as Table \ref{tab:performance_comparison}). These metrics are: (i) Audiobox-Aesthetics, assessing production quality (PQ), production complexity (PC), content enjoyment (CE), and content usefulness (CU); (ii) SongEval measuring coherence (Coh), memorability (Mem), natural vocal breathing/phrasing (NVBP), clarity of song structure (CSS), and overall musicality (OM).

As shown in Table~\ref{tab:objective_results}, SegTune shows consistent best performance with the smallest standard deviations.
As shown in Table~\ref{tab:objective_results}, SegTune shows consistent best performance with the smallest standard deviations. The proposed \textbf{SegTune-DPO} establishes a new state-of-the-art, ranking first in 8 out of 9 objective metrics, including CE and PQ. Although \textbf{SegTune-SFT} achieves the highest PC, SegTune-DPO offers a more balanced and high-quality generation, evidenced by its negligible variance compared to the fluctuating performance of models like LeVo and ACE-Step.

\section{Wilcoxon Signed-rank Test for Subjective Evaluation} 
\label{appendix:Wilcoxon}
To quantify the perceptual advantage of SegTune shown in figure \ref{fig:mos}, we conducted a Wilcoxon signed-rank test using expert Mean Opinion Scores (MOS). As detailed in Table~\ref{tab:wilcoxon_results}, SegTune demonstrates a statistically significant superiority in \textbf{Musicality} ($M = 4.57$) compared to all competing systems. The test statistics reveal decisive rejections of the null hypothesis ($p < .001$) against YuE ($M=3.07$), ACE-Step ($M=3.46$), DiffR.+ ($M=3.24$), and LeVo ($M=3.36$), with Wilcoxon values ($W$) clustering near zero, indicating a strong consensus among raters favoring SegTune.

In terms of \textbf{Quality}, SegTune ($M = 3.87$) maintains a significant edge over YuE ($M=2.58$, $W=18.0$), ACE-Step ($M=2.98$, $W=36.0$), and DiffR.+ ($M=3.10$, $W=13.5$), all yielding $p < .001$. Notably, the comparison against LeVo ($M=3.96$) resulted in a non-significant $p$-value of $0.524$ with a $W$ value of $164.0$. This suggests that while SegTune is generally preferred, its audio quality is statistically on par with the LeVo baseline under these specific evaluation conditions.

\begin{table}[!h]
  \centering
  \caption{Wilcoxon signed-rank test results comparing baseline models to SegTune}
  \label{tab:wilcoxon_results}
  \small
  \setlength{\tabcolsep}{3.5pt} 
  \begin{tabular}{@{}l l c c l@{}}
    \toprule
    \textbf{Metric (SegTune)} & \textbf{Baseline} & \textbf{Mean} & \textbf{$W$} & \textbf{$p$-value} \\
    \midrule
    \multirow{4}{*}{\makecell{Musicality \\ (4.57 $\pm$ 0.52)}} 
      & YuE      & 3.07 & 0.0   & $<.001^{***}$ \\
      & ACE-Step & 3.46 & 1.0   & $<.001^{***}$ \\
      & DiffR.+  & 3.24 & 11.0  & $<.001^{***}$ \\
      & LeVo     & 3.36 & 21.0  & $<.001^{***}$ \\
    \midrule
    \multirow{4}{*}{\makecell{Quality \\ (3.87 $\pm$ 0.56)}} 
      & YuE      & 2.58 & 18.0  & $<.001^{***}$ \\
      & ACE-Step & 2.98 & 36.0  & $<.001^{***}$ \\
      & DiffR.+  & 3.10 & 13.5  & $<.001^{***}$ \\
      & LeVo     & 3.96 & 164.0 & $0.524$       \\
    \bottomrule
  \end{tabular}
  
  \begin{minipage}{\linewidth}
    \footnotesize \textit{Note:} $W$ represents the Wilcoxon signed-rank value. 
  \end{minipage}
\end{table}

\section{Singer Gender Control Experiment and Visualization} 
\label{appendix:gender control}

To support a more interpretable and evidence-based comparison of the capacity of MuQ-MuLan and Qwen3-Embedding-0.6B in controlling vocal characteristics in song generation, we conduct controlled experiments and visualizations using \textit{singer gender} as a representative attribute. 

Specifically, we randomly sample 1,000 global prompts from the training dataset, each containing textual descriptions across multiple dimensions—including singer gender, musical style, and emotional tone. By applying rule-based string replacement, we invert the specified singer gender in each prompt (i.e., female vs. male), thereby constructing a paired control dataset of 2,000 prompts.
We then extract text embeddings for all 2,000 prompts using both MuQ-MuLan and Qwen3-Embedding, respectively. Each set of embeddings is first reduced to 50 dimensions via Principal Component Analysis and subsequently visualized using t-distributed Stochastic Neighbor Embedding (t-SNE). 

\begin{figure}[htp]
    \centering
    \includegraphics[width=\linewidth]{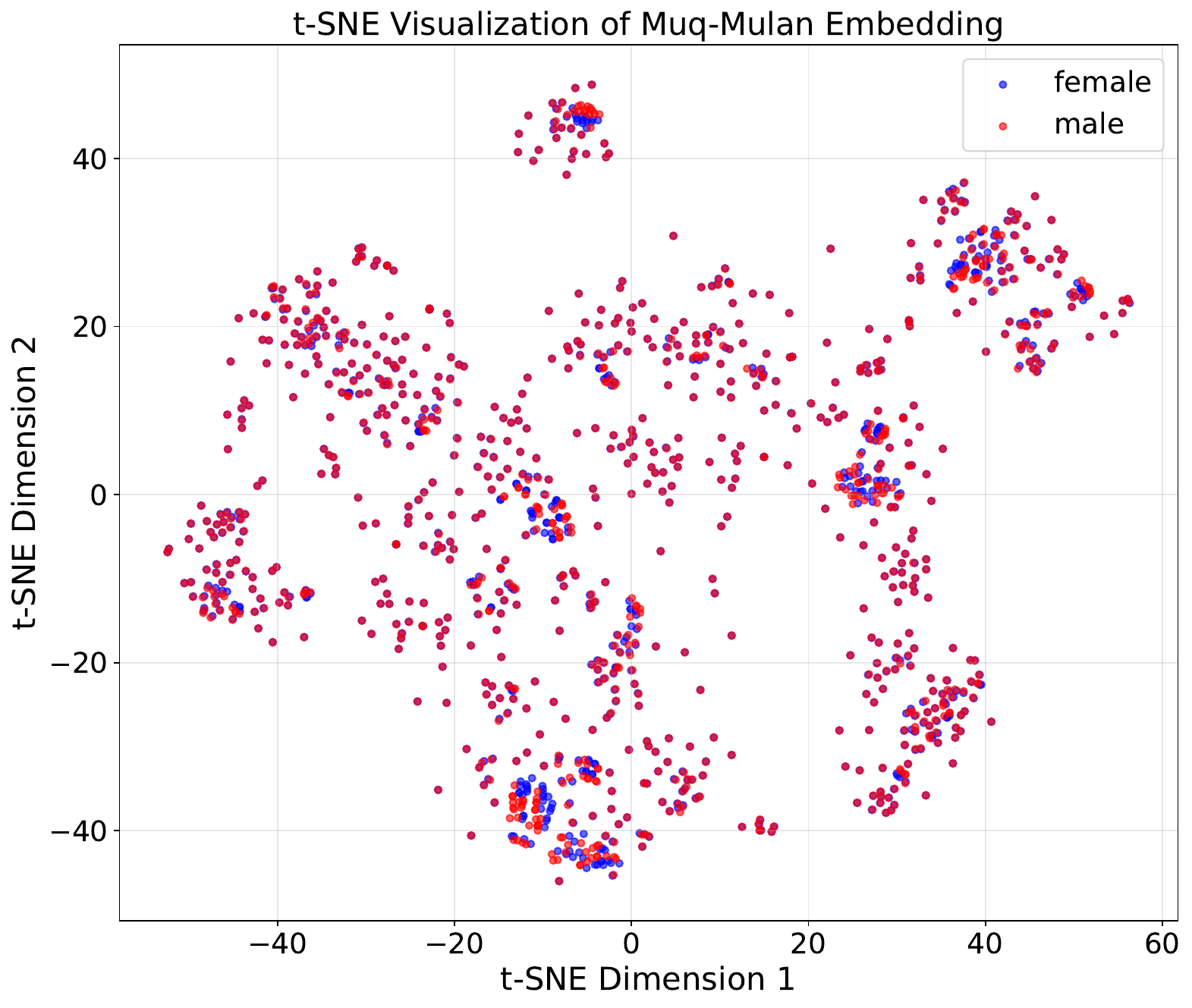}
    \caption{t-SNE visualization of Muq-Mulan embeddings on singer gender control.}
    
    \label{fig:mulan_vis}
\end{figure}

\begin{figure}[h]
    \centering
    \includegraphics[width=\linewidth]{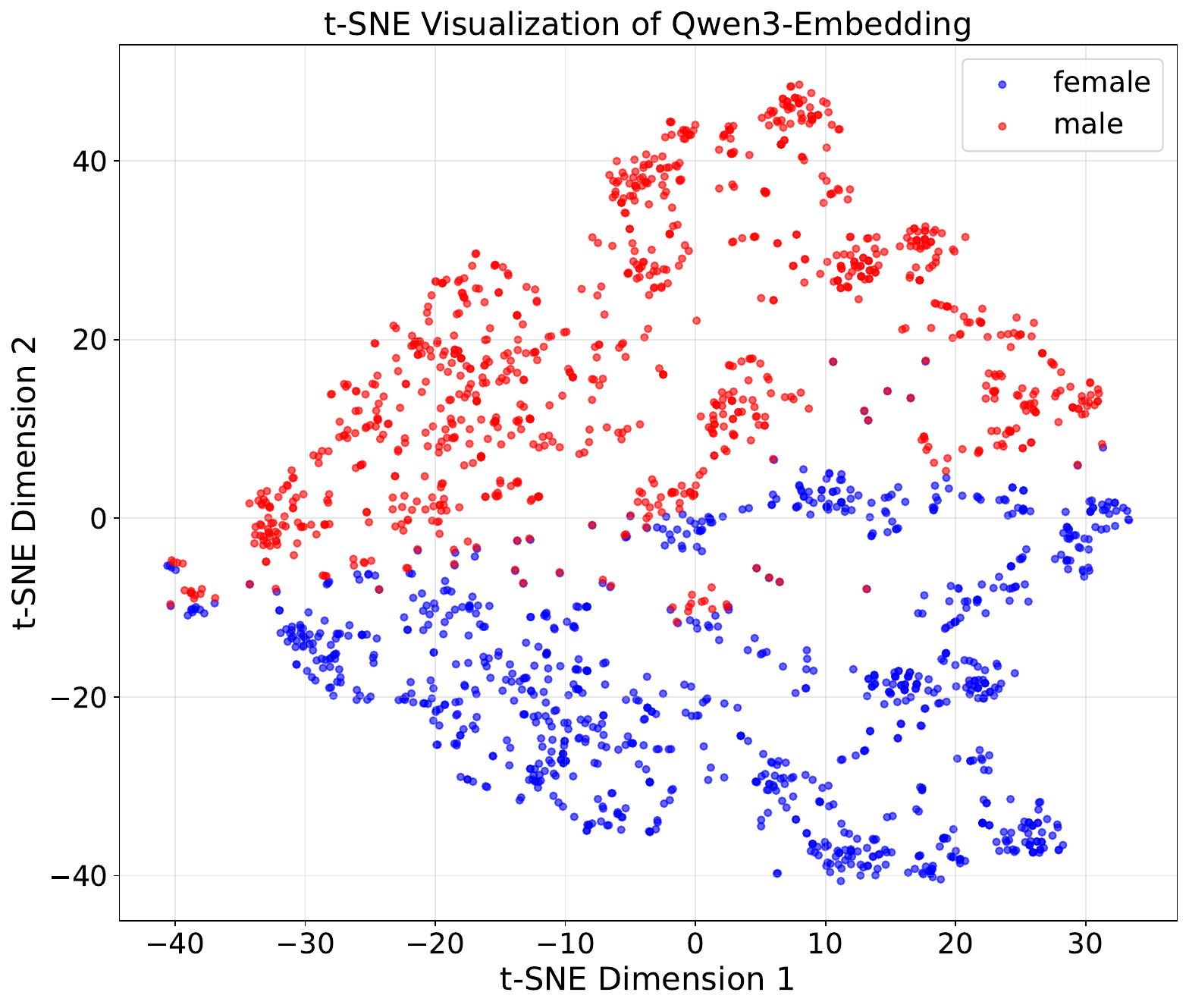}
    \caption{t-SNE visualization of Qwen3-Embedding on singer gender control.}
    
    \label{fig:qwen3_vis}
\end{figure}

As shown in the figure \ref{fig:mulan_vis} and figure \ref{fig:qwen3_vis}, the t-SNE visualization reveals a stark contrast: embeddings from MuQ-MuLan exhibit near-complete overlap between female and male prompts in the embedding space, indicating an inability to encode gender-related vocal attributes, and the cosine distance of cluster centers for female and male groups is 0.002.
In contrast, Qwen3-Embedding yields clearly separated clusters for the two genders, and the cosine distance of cluster centers for female and male groups is 0.107, providing direct empirical evidence of its superior discriminative capability.

\end{document}